\documentclass[doublecol,final]{epl2} 

\usepackage[]{graphicx}
\usepackage{hyperref}
\usepackage{amsmath, amssymb}
\usepackage{subfigure}
\usepackage{lmodern}
\usepackage{xcolor}

\newcommand{\invsqcm}{cm\ensuremath{^{-2}}}
\newcommand{\beq}{\begin{equation}\begin{aligned}}
\newcommand{\eeq}{\end{aligned}\end{equation}}

\newcommand{\half}{\ensuremath{\frac{1}{2}\,}}
\renewcommand{\vect}[1]{\ensuremath{\boldsymbol{#1}}\,}
\newcommand{\sio}{SiO$_2\,$}
\newcommand{\bra}[1]{\ensuremath{\langle #1 |\,}}
\newcommand{\ket}[1]{\ensuremath{|#1\rangle\,}}
\renewcommand{\d}{\,\textmd{d}}
\newcommand{\volt}{\textrm{V}\,}

\definecolor{linkcol}{rgb}{0,0,0.4} 
\definecolor{citecol}{rgb}{0.5,0,0} 

\hypersetup{
bookmarksopen=true,
pdftitle="Infrared spectroscopy of hole doped ABA-stacked trilayer graphene",
pdfauthor="N. Ubrig et al.", 
pdfsubject="Infrared spectroscopy of hole doped ABA-stacked trilayer graphene", 
pdfstartview={FitH},    
pdfmenubar=true, 
pdfhighlight=/O, 
colorlinks=true, 
pdfpagemode=UseNone, 
pdfpagelayout=SinglePage, 
pdffitwindow=true, 
linkcolor=linkcol, 
citecolor=linkcol, 
urlcolor=blue
}

\title{Infrared spectroscopy of hole doped ABA-stacked trilayer graphene}
\subtitle{}
\author{N. Ubrig\inst{1} \and P. Blake\inst{2} \and D. van der Marel\inst{1} \and A.B. Kuzmenko\inst{1}\thanks{E-mail: \email{alexey.kuzmenko@unige.ch}}
   }
\shortauthor{N. Ubrig et al.}

\institute{\inst{1} D\'epartement de Physique de la Mati\`ere Condens\'ee , Universit\'e de Gen\`eve, CH-1211 Gen\`eve 4, Switzerland\\
	  \inst{2} Graphene Industries Ltd, Manchester Centre for Mesoscience and Nanotechnology, University of Manchester, Manchester, M13 9PL, United Kingdom}


\abstract{
Using infrared spectroscopy, we investigate bottom gated ABA-stacked trilayer graphene subject to an additional environment-induced p-type doping. We find that the Slonczewski-Weiss-McClure tight-binding model and the Kubo formula reproduce the gate voltage-modulated reflectivity spectra very accurately. This allows us to determine the charge densities and the potentials of the $\pi$-band electrons on all graphene layers separately and to extract the interlayer permittivity due to higher energy bands.}

\begin{document}
\maketitle%
\section{Introduction}
\label{intro}

Trilayer graphene has recently attracted much interest because its electronic structure is distinctly different from the bands found in more studied monolayer and bilayer graphene. Interestingly, the two major stacking types of trilayer graphene, namely the Bernal (or ABA-)stacking and the rhombohedral (ABC-)stacking, result in rather different band properties \cite{guinea_electronic_2006}. In rhombohedral graphene, a sizeable bandgap can be generated and tuned with an external electric field \cite{aoki_dependence_2007,zhang_band_2010,lui_observation_2011,avetisyan_stacking_2010}, while in Bernal stacked graphene the dominant effect is gate-tunable band overlap \cite{aoki_dependence_2007,lu_influence_2006,koshino_gate-induced_2009,avetisyan_electric_2009,craciun_trilayer_2009,bao_stacking-dependent_2011,henriksen_quantum_2012}, resulting in a semimetallic behaviour. Trilayer graphene presents the simplest case where the complete set of Slon\-cze\-wski-Weiss-McClure (SWM) tight-binding parameters are needed 
to describe the band structure, as it is the case for three-dimensional graphite \cite{mcclure_band_1957,slonczewski_band_1958}. However, in contrast to graphite, trilayer graphene allows exploring different doping regimes and the effect of an electric field on the band structure. Thus it might be beneficial for understanding the complex graphite physics to study its trilayer counterpart.

\begin{figure}
 \centering
  \subfigure[]{\label{fig:setup}\includegraphics[width=.235\textwidth]{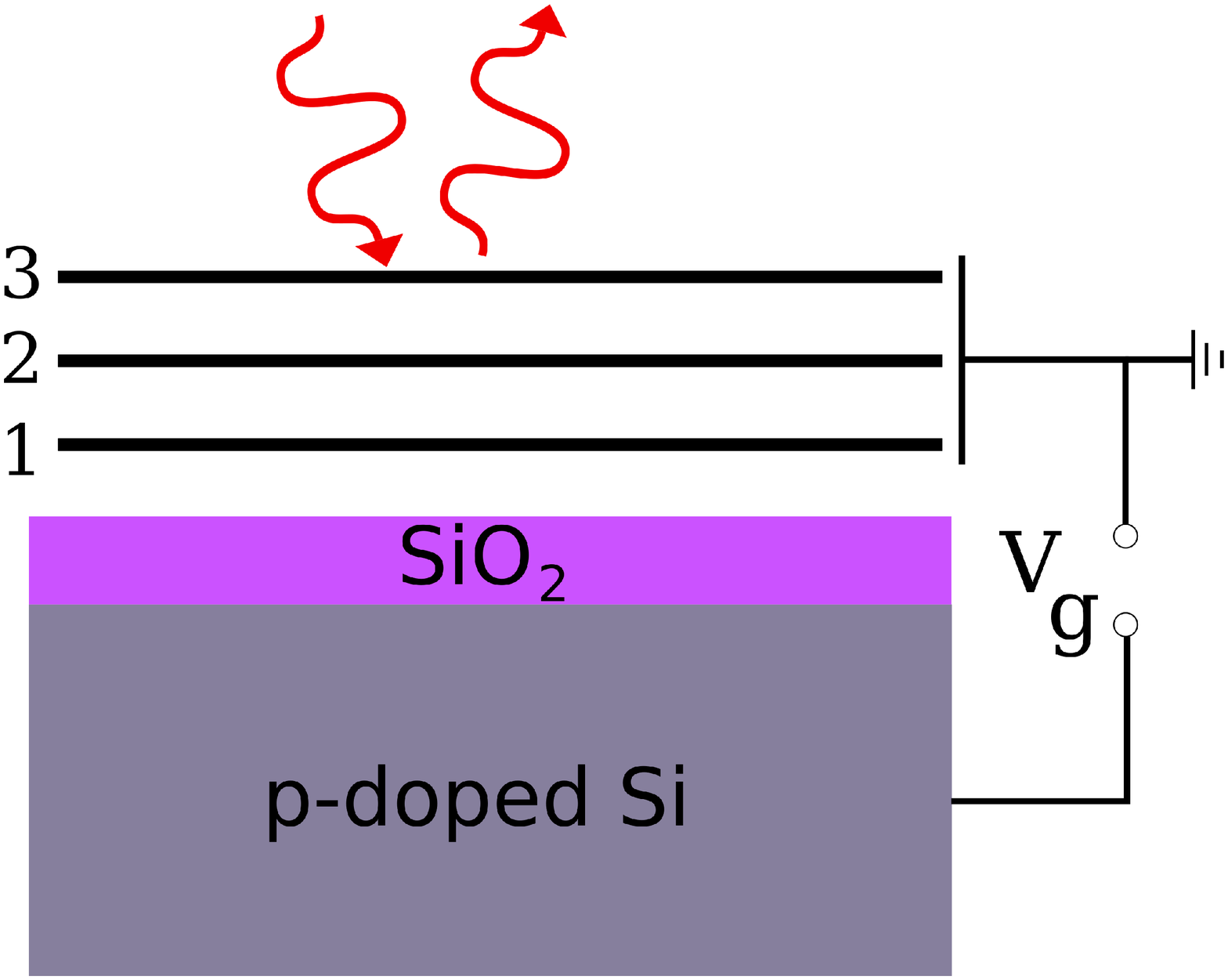}}\hfill%
  \subfigure[]{\label{fig:meanR}\includegraphics[width=.235\textwidth]{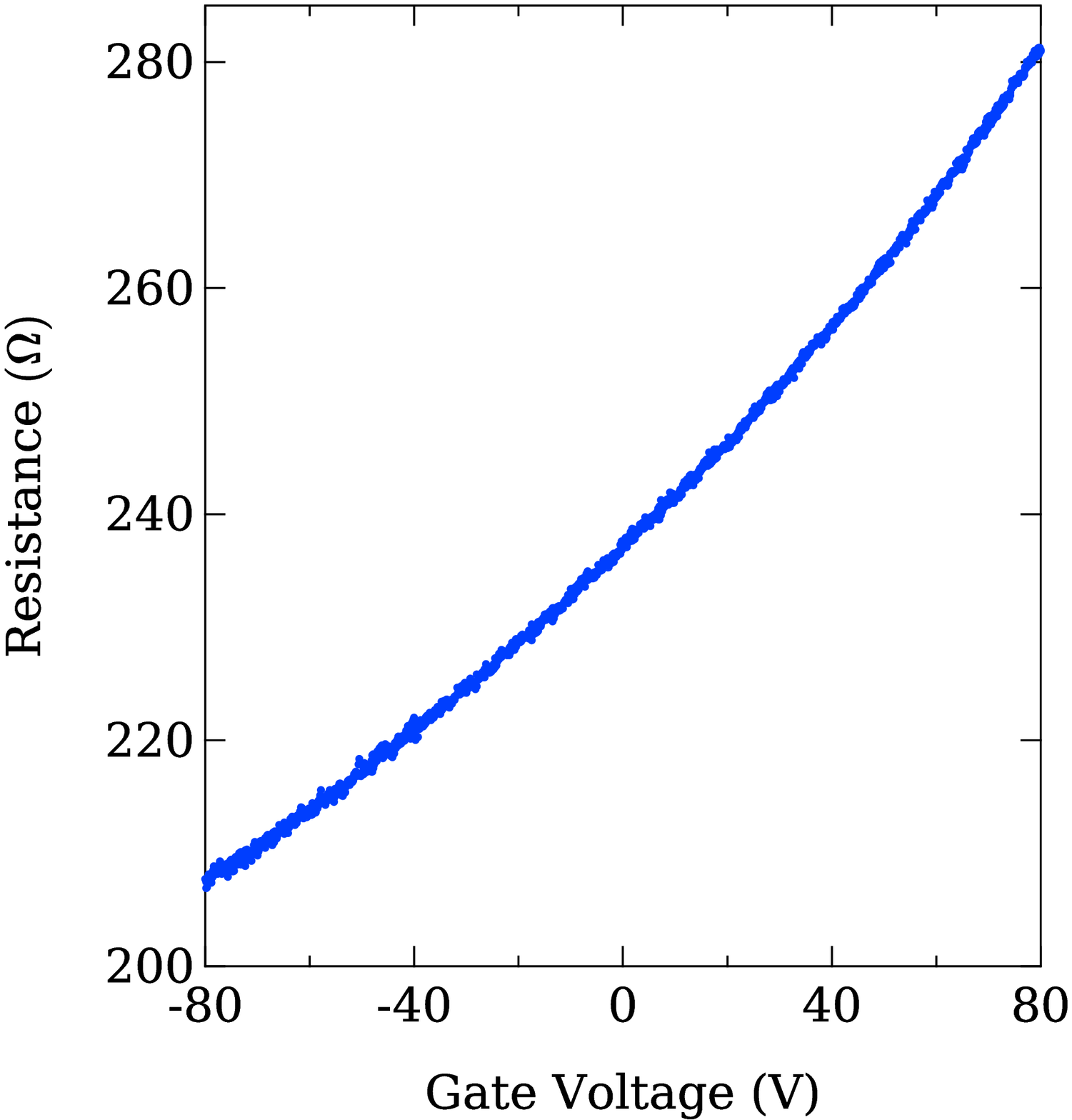}}%
  \caption{\subref{fig:setup} Schematic representation of bottom gated three-layer graphene used in this study. The layer numbering used in the text is given. \subref{fig:meanR} The resistance of the sample as a function of the gate voltage at 20 K.}
  \label{fig:RandSetup}
\end{figure}

In this work we focus on the ABA-stacked trilayer graphene, the form most often found in nature, which is plane symmetric with respect to the middle layer. Our goal is to determine experimentally optical properties as a function of an external electric field and check the ability of the existing theories to reproduce them quantitatively. The sample is a bottom gated exfoliated flake. Its charge neutrality point is strongly shifted with respect to zero gate voltage, which allows us to explore the high doping regime not achieved so far in a combination with infrared spectroscopy. By treating the electrostatic potentials of each of the three layers as gate voltage-dependent free parameters, we achieve a very good match between the experiment and the tight-binding model. With the help of a basic electrostatic model for trilayer graphene \cite{koshino_gate-induced_2009,avetisyan_electric_2009} we find an accurate value for the static perpendicular interplane permittivity $\varepsilon_r$ due to the cumulative 
polarizability of all electronic shells except the low-energy $\pi$-bands, considered explicitly by the tight-binding model.

\section{Experimental results}
\label{sec:experiment}

A large, several tens of microns, flake of trilayer graphene was obtained by mechanical exfoliation \cite{novoselov_two-dimensional_2005} and deposited on a Si substrate covered by a 300 nm layer of silicon oxide (Fig. \ref{fig:setup}). Electrical contacts were deposited by evaporating gold through a shadow mask. The stacking was identified as ABA-type, based on Raman spectroscopy \cite{lui_imaging_2010} and infrared spectroscopy as discussed below.\\
The resistivity curve (Fig. \ref{fig:meanR}) shows that the sample is strongly hole doped so that the charge neutrality point cannot be achieved even by applying $V_g = $ +80 V to the gate. Annealing the sample several times at 200 $^{o}$C in a H$_2$ - N$_2$ atmosphere did not change significantly the resistivity curve. We therefore attribute the extrinsic doping either to gold atoms, generally known to dope graphene positively \cite{gierz_atomic_2008,klusek_graphene_2009}, scattered below the shadow mask onto the graphene surface during the contact evaporation or to a layer of water or hydrocarbon contamination trapped below the flake \cite{haigh_cross-sectional_2012}. Although not being able to reach the electron-type doping, we benefited from this situation by measuring in the regime of high hole doping ($\sim 3\cdot 10^{13}$ \invsqcm), which would not be achievable with an initially undoped sample.\\
The reflectivity measurements at various gate voltages between -80 V and +80 V with a step of 20 V were performed at 20 K in a helium-flow cryostat
mounted on an infrared microscope coupled to a Fourier transform infrared spectrometer. The infrared spot of about 15$\times$15 $\mu$m was kept far from the flake edges and electrical contacts. Each measurement on graphene was followed by recording reference spectra on the bare substrate and on gold. A black-body Globar light source was used between 0.1 and 0.7 eV and a tungsten lamp was employed between 0.4 and 1.5 eV.

\begin{figure}
 \centering
  \includegraphics[width=.48\textwidth]{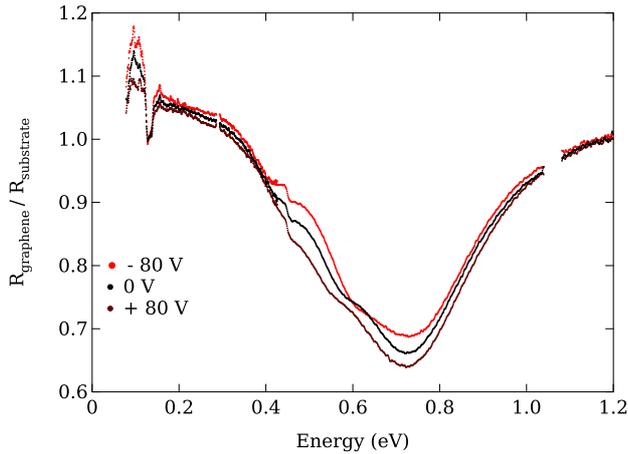}
  \caption{Reflectivity of trilayer graphene at 20 K for different gate voltages of -80, 0 and +80 V.}
  \label{fig:RgRs}
\end{figure}

The reflectivity of trilayer graphene normalized by the bare substrate is plotted in Fig. \ref{fig:RgRs} for gate voltages of -80, 0 and +80 V. Although the changes between the spectra are caused by the modification of the optical properties of graphene, some profound spectral features present in all spectra are due to the substrate. For example, the strong structure at about 0.1 eV is due to optical phonons in \sio. The deep minimum at 0.7 eV, where the presence of graphene reduces the reflected signal by almost 40 percent, is due to the Fabry-P\'{e}rot interference in the oxide layer.\\
Even though the substrate somewhat complicates the interpretation of the measured spectra, some qualitative observations can be made without data modeling. First, at energies below about 1 eV the reflectivity systematically decreases as the gate voltage increases. This is directly related to the decreasing metallicity and Drude weight as the concentration of holes is reduced. Second, above 1 eV the curves essentially overlay showing that the electric field has little or no influence on the optical properties at high energies. Third, in the region between about 0.4 and 0.8 eV, a more complex gate voltage dependence is observed, which is due to interband transitions in graphene.

\begin{figure}
\centering
\includegraphics[width=\columnwidth]{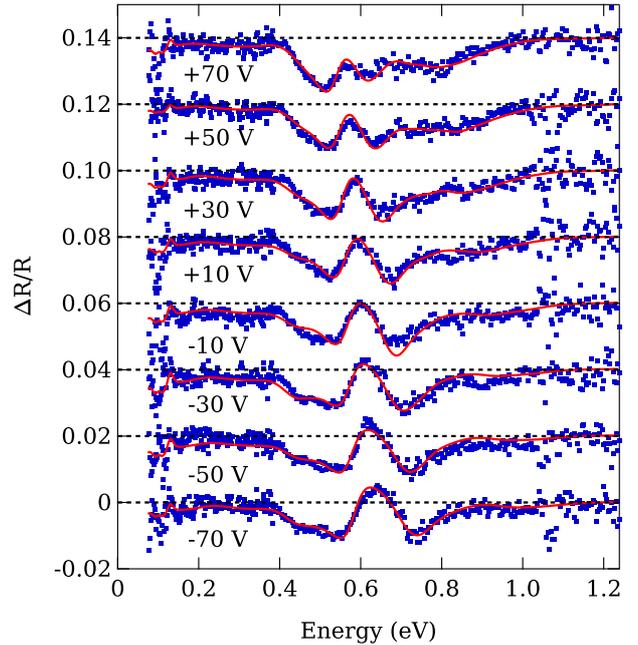}%
\caption{Experimental differential reflectivity for pairs of consecutive gate voltages. The value given for each curve is the average of the two voltages (separated by 20 V) used to generate the difference. For clarity each curve is shifted and the zero levels are indicated by dotted lines. The points represent the experimental data while the solid red line is the result of the fit.}
\label{fig:point2_DR_R}
\end{figure}

In order to emphasize the effect of the gate and to extract more information from the data, we show in Fig. \ref{fig:point2_DR_R} the differential (or contrast) spectra for all pairs of consecutive gate voltages. They are defined as follows \cite{kuzmenko_determination_2009}:

\beq
\frac{\Delta R}{R}(\omega, V_g) = 2 \frac{R(\omega, V_{g} + 10V) - R(\omega, V_{g} - 10V)}{R(\omega, V_{g} + 10 V) + R(\omega, V_{g} - 10 V)}.
\label{eq:contrast}
\eeq

\noindent The curves are marked by pronounced spectral structures showing a strong gate dependence. For example, the position of the peak at around 0.6 eV shifts by almost 0.1 eV as the gate voltage is swept from +70 V to -70 V. The shape of the structures is also strongly influenced by the gate. Such a non-trivial behavior of the contrast spectra is due to the fact that not only the chemical potential is modified by the gate voltage but also the band structure is affected by the gate-induced electric field. Thus quantitative modeling is required in order to identify spectral features and to get further insight into the effect of the gate on the electronic structure of the sample. Below we discuss this analysis in detail. Looking ahead, we note here that a very good agreement between the data and the tight-binding model of the ABA-stacked graphene is achieved (red solid lines in Fig. \ref{fig:point2_DR_R}). Given that the band structures of ABC- and ABA-stacked graphene are entirely different\cite{guinea_electronic_2006}, we conclude that the sample is ABA-stacked.

\section{Optical data modeling}
\label{sec:theory}

\begin{figure}
\centering
\includegraphics[width=.35\textwidth]{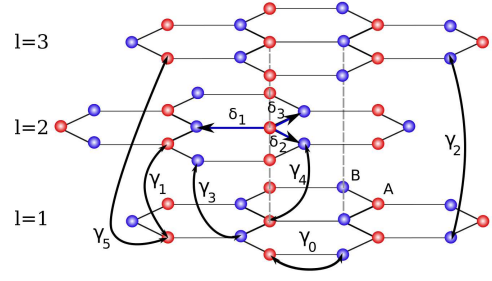}
\caption{The crystal lattice of ABA-stacked trilayer graphene with the definition of the hopping parameters.}
\label{fig:trilayerTB}
\end{figure}

To model the optical data we use a tight-binding description of trilayer graphene \cite{koshino_gate-induced_2009,avetisyan_electric_2009} based on the SWM-model of graphite \cite{mcclure_band_1957,slonczewski_band_1958}. The hopping terms $\gamma_0$, $\gamma_1$, $\gamma_2$, $\gamma_3$, $\gamma_4$ and $\gamma_5$ are illustrated in Fig. \ref{fig:trilayerTB}. Additionally, $\Delta$ represents the difference between the on-site energy for the two sublattices A and B (marked with red and blue colors respectively). To account for the effect of an external electric field we add potentials $U_1$, $U_2$ and $U_3$ to the first, second and the third layers respectively \cite{koshino_gate-induced_2009}. In the basis of \ket{A_1}, \ket{B_1}, \ket{A_2}, \ket{B_2}, \ket{A_3} and \ket{B_3} the Hamiltonian of ABA-stacked trilayer graphene reads:

\beq
\mathbf{H} = {\footnotesize
\left(\begin{array}{cccccc}
		      U_1	&\gamma_0 \phi	&-\gamma_4 \phi	&\gamma_3 \phi^*	&\gamma_2	&0\\
		      \gamma_0 \phi^*	&U_1 + \Delta			&\gamma_1	&-\gamma_4 \phi	&0	&\gamma_5\\
		      -\gamma_4 \phi^*	&\gamma_1			&U_2 + \Delta	&\gamma_0 \phi	&-\gamma_4 \phi^*	&\gamma_1\\
		      \gamma_3 \phi	&-\gamma_4 \phi^*	&\gamma_0 \phi^*	&U_2	&\gamma_3 \phi	&-\gamma_4 \phi^*\\
		      \gamma_2				&0			&-\gamma_4 \phi	&\gamma_3 \phi^*	&U_3	&\gamma_0 \phi\\
		      0					&\gamma_5			&\gamma_1	&-\gamma_4 \phi	&\gamma_0 \phi^*	&U_3 + \Delta
                   \end{array}
             \right) }
\label{eq:hamiltonian}
\eeq

with

\beq
\phi = e^{i\vect{k}\vect{\delta}_1} + e^{i\vect{k}\vect{\delta}_2} + e^{i\vect{k}\vect{\delta}_3}
\label{eq:phi}
\eeq

where $\vect{k}$ is the momentum with respect to the K-point of the Brillouin zone and the vectors \vect{\delta}$_{1,2,3}$ connect the nearest neighbor atoms as shown in Fig. \ref{fig:trilayerTB}.\\
In order to obtain the optical properties of this system, within the linear response theory, we use the Kubo formula for the optical sheet conductivity $\sigma(\omega)$. We calculate the Drude and the interband terms using the following expressions

\beq
\sigma_D = \frac{2 \sigma_0}{\pi} \sum_i &\int \d^2 \vect{k} \left| \bra{\vect{k},i} \frac{\partial \mathbf{H}}{\partial k_x} \ket{\vect{k},i}  \right|^2 \\
			       &\cdot \frac{-\partial f(\epsilon_{\vect{k},i})}{\partial \epsilon}\cdot \frac{i}{\hbar\omega + i\Gamma_D}
\label{eq:kubo_drude}
\eeq

and

\beq
\sigma_{IB} = &\frac{2 \sigma_0}{\pi} \sum_{i,j\neq i} \int \d^2 \vect{k} \left| \bra{\vect{k},i} \frac{\partial \mathbf{H}}{\partial k_x} \ket{\vect{k},j}  \right|^2 \\
				         &\cdot \frac{f(\epsilon_{\vect{k},i}) - f(\epsilon_{\vect{k},j})}{\epsilon_{\vect{k},i} - \epsilon_{\vect{k},j}} \cdot \frac{i}{\hbar\omega - \epsilon_{\vect{k},j} + \epsilon_{\vect{k},i} + i\Gamma},
\label{eq:kubo_IB}
\eeq

where $\sigma_0 = e^2/4\hbar \approx 6.08\cdot 10^{-5}$ $\Omega^{-1}$ is the universal AC conductance of monolayer graphene \cite{ando_dynamical_2002}, $\epsilon_{\vect{k},i}$ is the band energy ($i$ = 1,..,6 ), $f(\epsilon) =  \left( 1 + {\rm exp}\left( \frac{\epsilon - \mu}{k_B T} \right) \right)^{-1}$ is the Fermi-Dirac distribution and $\mu$ is the chemical potential. $\Gamma$ and $\Gamma_D$ are the electronic broadening parameters for the interband and the intraband transitions respectively. For simplicity, the interband broadening is taken to be momentum-, energy-, band- and doping-independent.

The total charge density $n_{\rm total}$ and the charge densities of each layer, $n_l$ ($l = 1,2,3$), can also be calculated through

\beq
n_{\rm total} = \frac{1}{2\pi^2} \sum_i \int \d^2\vect{k}\left( f(\epsilon_{i,\vect{k}}) - \half \right)
\label{eq:ntotal}
\eeq

and

\beq
n_l = \frac{1}{2\pi^2} &\sum_i \int \d^2\vect{k}\left( f(\epsilon_{i,\vect{k}}) - \half \right)\\
		       &\cdot \left( |\psi_{A_l}|^2 + | \psi_{B_l}|^2 \right),
\label{eq:nlayer}
\eeq

where we subtract \half  in order to count the doping level with respect to half filling of the $\pi$ bands. $\psi_{A_l}$ and $\psi_{B_l}$ are the projections of the total electronic eigenfunction on the sub-lattices $A_l$ and $B_l$.\\
Once the conductivity is determined we can calculate the optical reflectivity of the complete graphene/\sio/Si  system using Fresnel eq.s and consequently compute the differential reflectivity (Equation \ref{eq:contrast}). As $\Delta R/ R$ values are rather small it is useful to introduce the so-called sensitivity functions \cite{kuzmenko_determination_2009} $\beta_1(\omega)$, $\beta_2(\omega)$, which are determined by the substrate, and to employ the linearized formula

\beq
\frac{\Delta R}{R} \approx \beta_1(\omega) \frac{\textrm{Re} \Delta \sigma(\omega)}{\sigma_0} + \beta_2(\omega) \frac{\textrm{Im} \Delta \sigma(\omega)}{\sigma_0}
\eeq

When fitting the experimental data, we fixed the SWM parameters to the values published in Ref. \cite{partoens_graphene_2006}, which we reproduce in Table \ref{tab:TBparameters} for the sake of convenience. We also fixed the Drude broadening $\Gamma_{D}$ to 5 meV \cite{kuzmenko_universal_2008} as its value has almost no influence on the optical data in the studied spectral range. In contrast, the interband broadening $\Gamma$ strongly affects the data, resulting in the overall smoothening of the spectral structures. It was therefore taken as an adjustable parameter, for which the best value was found to be 45 meV.\\
The most important detail of our fitting procedure is that we treat the chemical potential and the plane potentials $U_l$ as independent adjustable parameters at every value of $V_g$, in order to extract the actual gate voltage dependence of these quantities from the optical data. It is easy to see that shifting the values of $\mu$, $U_1$, $U_2$ and $U_3$ by the same amount does not change any physical properties such as the optical spectra and the charge density. To remove this ambiguity, we imposed an additional condition $U_1 + U_2 + U_3 = 0$.\\
The best fits are shown in Fig. \ref{fig:point2_DR_R} with the solid red lines. One can notice an excellent match between the experiment and the model, including all spectral features at all gate voltages. It is worth to mention that although the differential reflectivity between 0.15 and 0.4 eV is rather featureless it is distinctly non-zero. This is caused by the gate variable Drude contribution and the model reproduces this fact very well.\\
As all the curves were fitted \emph{simultaneously}, the obtained parameters were well defined and converged to the same numbers,  which we carefully tested by using different fitting procedures and by varying the initial parameter values. The robustness of these results allows us to develop a deeper analysis of the response of graphene to the external electric field which we present in the next section.

\begin{table}
\caption{SWM band parameters used in this work to fit the optical spectra. The values are taken from Ref. \cite{partoens_graphene_2006}.}
\label{tab:TBparameters}       
\centering
\begin{tabular}{lcr}
\hline\noalign{\smallskip}
Parameter &\hspace{.3\columnwidth}	&Value (eV)\\
\noalign{\smallskip}\hline\noalign{\smallskip}
$\gamma_0$ &\hspace{.3\columnwidth}	&3.12\\
$\gamma_1$ &\hspace{.3\columnwidth}	&0.377\\
$\gamma_2$ &\hspace{.3\columnwidth}	&-0.0206\\
$\gamma_3$ &\hspace{.3\columnwidth}	&0.29\\
$\gamma_4$ &\hspace{.3\columnwidth}	&0.12\\
$\gamma_5$ &\hspace{.3\columnwidth}	&0.025\\
$\Delta$ &\hspace{.3\columnwidth}	&0.0138\\
\noalign{\smallskip}\hline
\end{tabular}
\end{table}

\section{Discussion}
\label{sec:discuss}

\begin{figure*}
 \centering
  \subfigure[]{\label{fig:p2_S1}\includegraphics[width=.32\textwidth]{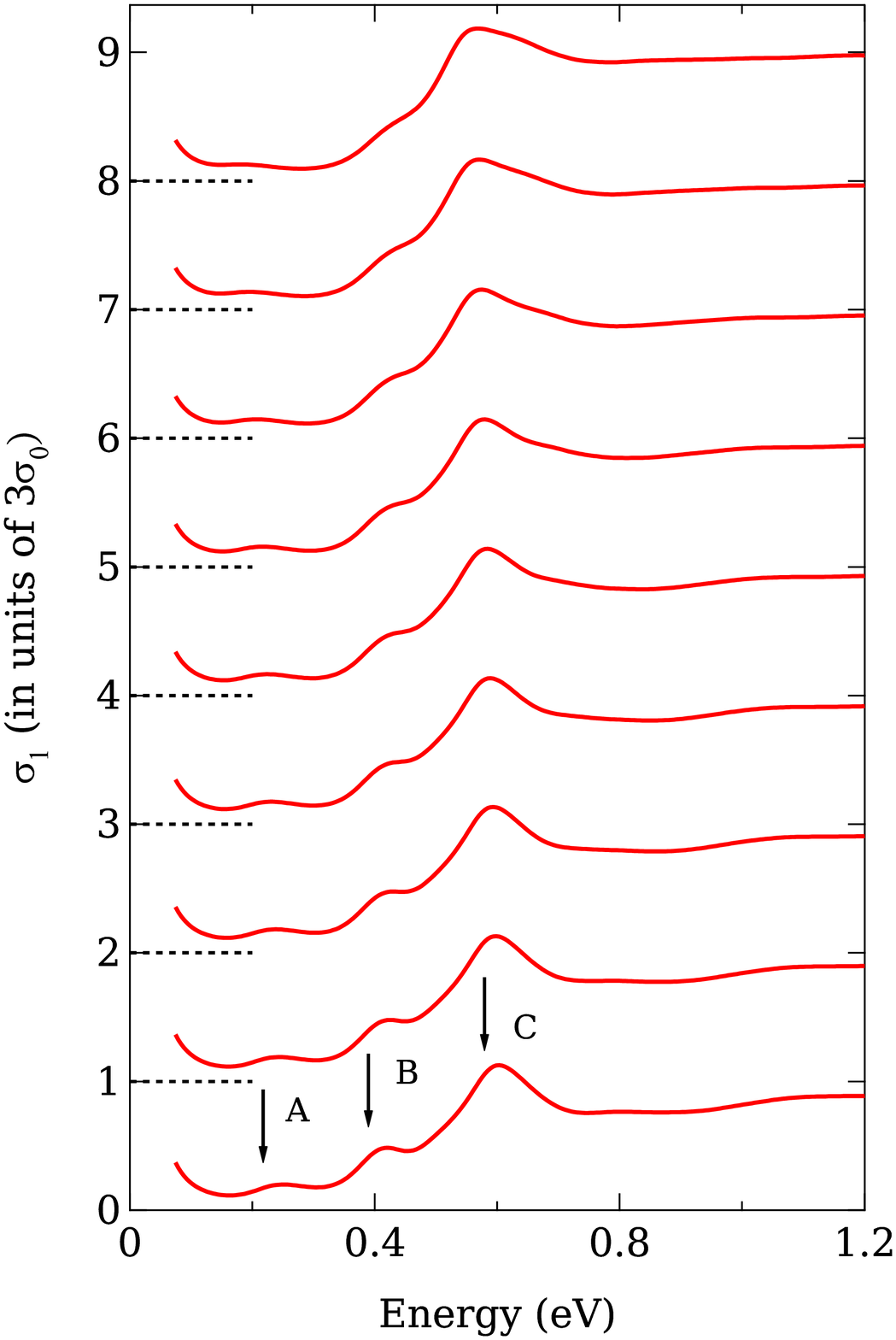}}%
  \subfigure[]{\label{fig:fitparam}\includegraphics[width=.32\textwidth]{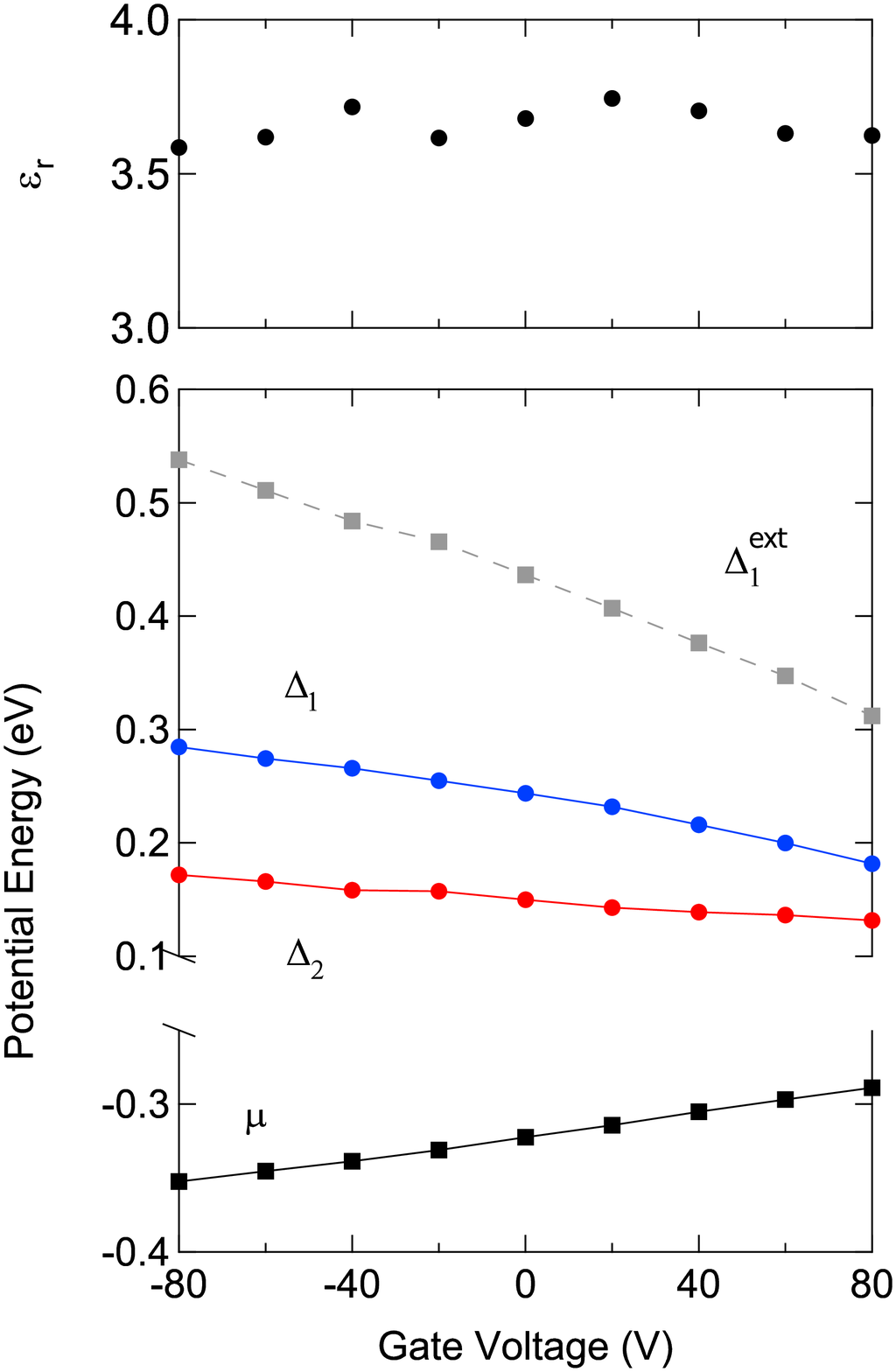}}%
  \subfigure[]{\label{fig:nvsgate}\includegraphics[width=.32\textwidth]{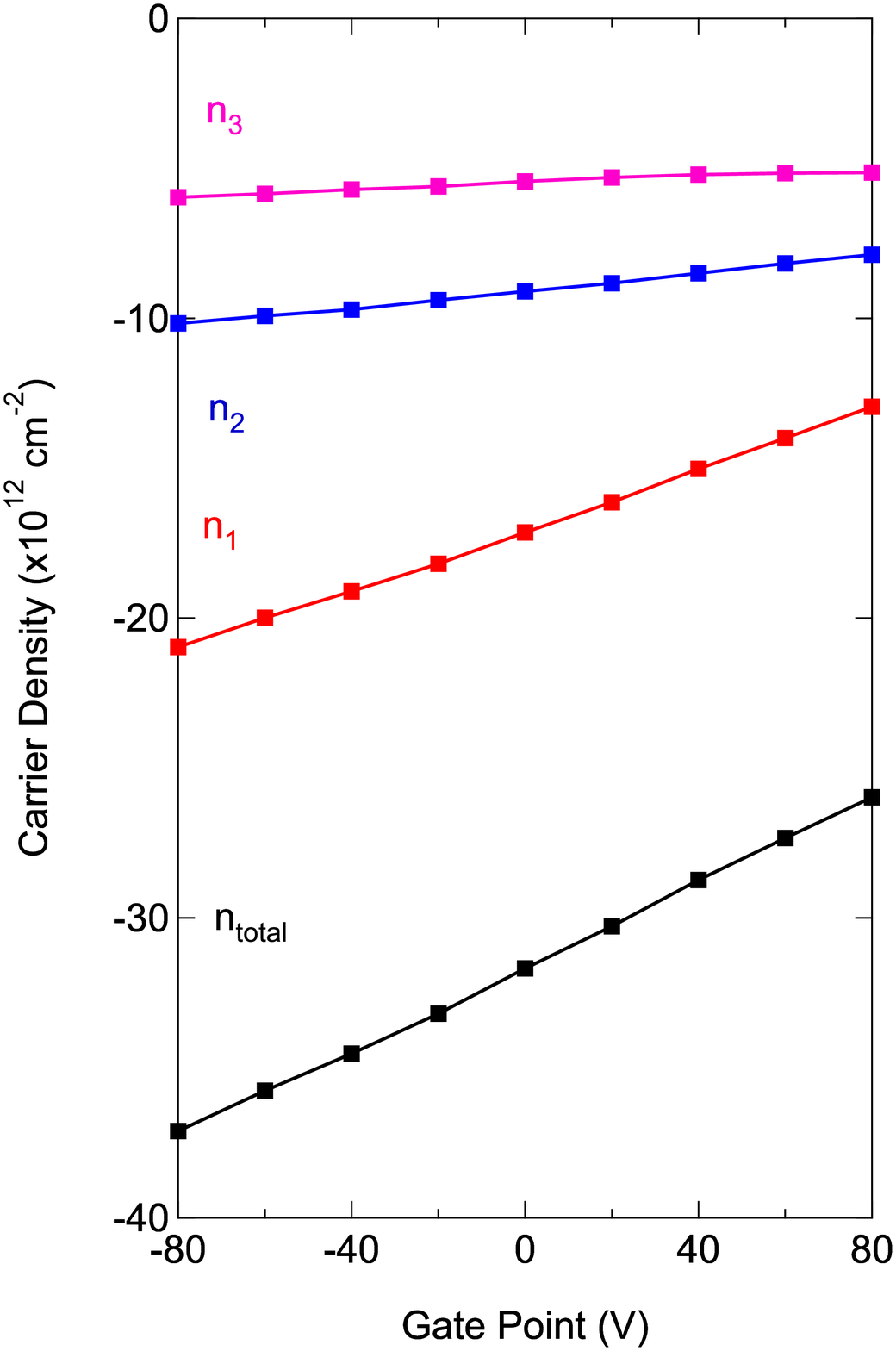}}%
\caption{\subref{fig:p2_S1} The calculated optical sheet conductivity at different gate voltages obtained from the fits shown in Fig. \ref{fig:point2_DR_R} as described in the text. The arrows mark the dominant peaks corresponding to the interband transitions indicated in Fig. \ref{fig:bands}. \subref{fig:fitparam} The interplane potentials $\Delta_1$, $\Delta_2$ and the chemical potential $\mu$ as a function of the gate voltage, directly obtained from the optical fits. Also shown are the derived values of the external potential $\Delta^{\rm ext}_1$ and the interspace permittivity $\varepsilon_r$. \subref{fig:nvsgate} The charge densities $n_l$ of each layer and the total density $n_{\rm total}$  as a function of the gate voltage.}
\label{fig:fitparams}
\end{figure*}

\begin{figure}[h!]
 \centering
  \includegraphics[width=.45\textwidth]{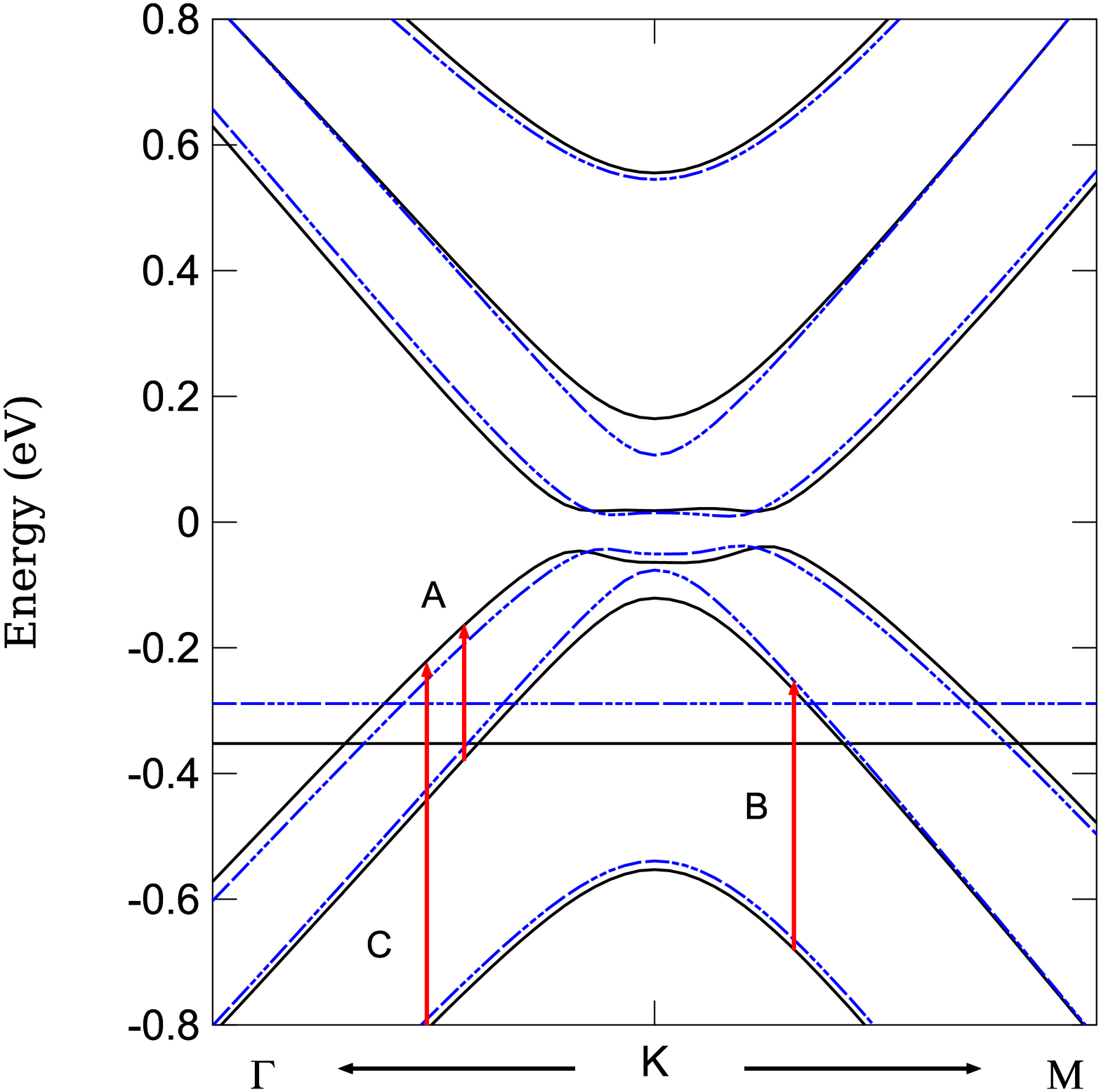}%
\caption{Electronic bands at $V_g$ = -80 V (solid black lines) and + 80V (dashed blue lines) calculated based on the fits in Fig. \ref{fig:point2_DR_R}. The horizontal lines indicate the chemical potential. The arrows correspond to the optical peaks indicated in Fig. \ref{fig:p2_S1}.}
\label{fig:bands}
\end{figure}

The optical conductivity spectra corresponding to the fits described above are shown in Fig. \ref{fig:p2_S1} for each value of the gate voltage. For illustration purposes, the curves are normalized by $3\cdot\sigma_{0}$. At low photon energies but above the Drude peak, the optical conductivity is significantly below $3\cdot\sigma_{0}$, while at high photon energies it approaches this value. This is related to the Pauli blocking, which results, in the case of monolayer graphene, in a conductivity step at 2$\mu$ \cite{li_dirac_2008}. Additionally, significant spectral features are the peaks at approximately 0.25 eV, 0.42 eV and 0.6 eV, which are denoted respectively as A, B and C in Fig. \ref{fig:p2_S1}. They can be assigned to transitions between almost parallel bands, which are illustrated in Fig. \ref{fig:bands}. The energy of the strongest peak (C) is close to $\sqrt{2}\gamma_1$, in agreement with the theoretical predictions \cite{lu_influence_2006,koshino_magneto-optical_2008,koshino_electronic_2009} and 
experimental data \cite{mak_evolution_2010,lui_imaging_2010,lui_observation_2011,henriksen_quantum_2012}. The peaks noticeably shift as a function of the gate voltage. For example, peak C moves from 0.55 eV at $V_g$ = +80 V to 0.6 eV at $V_g$ = -80 V. As can be seen in Fig. \ref{fig:bands}, this effect stems from the field-induced displacement of the bands with respect to each other. The transitions are broad because the scattering parameter $\Gamma$ is large, as compared to the numbers obtained in low doped samples \cite{lui_observation_2011}. We believe that extra scattering present here is due to the charged impurities that also give rises to the elevated doping.

Let us now inspect the dependence of the doping and the interplane potentials on the gate voltage. In Fig. \ref{fig:fitparam} we plot the chemical potential as a function of the applied gate. It is always negative with a high absolute value and moves gradually towards the charge neutrality as the gate voltage is increased. This corroborates our initial observation of a high level of hole doping.\\
The total charge carrier density $n_{\rm total}$ obtained using Equation \ref{eq:ntotal} is shown in Fig. \ref{fig:nvsgate}. At zero voltage the concentration of holes is about $3\cdot 10^{13}$ \invsqcm. It changes linearly with $V_g$ with a slope $\d n_{\rm total} / \d V_g$ equal to 7.0 $\cdot$ 10$^{10}$ \invsqcm \volt$^{-1}$. Remarkably, this value, which is based on the optical data only, agrees extremely well with the expected slope of 7.2 $\cdot$ 10$^{10}$ \invsqcm \volt$^{-1}$ for the actual thickness of the oxide layer \cite{novoselov_two-dimensional_2005}, thus providing extra credence to our analysis. From this dependence, one expects to reach the charge neutrality point at $V_g\approx$ +450 V, which is not achievable in the present experiment.\\
In the same Fig., we present the charge density on each layer separately according to Equation \ref{eq:nlayer}. As expected, the layer closest to the gate shows the strongest charge variation as a function of the external field, while the effect of the gate decreases strongly with increasing layer index due to internal screening. This is in good agreement with theoretical predictions \cite{avetisyan_electric_2009}.\\
In Fig. \ref{fig:fitparam} we plot the potential difference between the outer layers, $\Delta_1 = U_1 - U_3$, and twice the difference between the average of the potentials of outer layers and the one of the central layer, $\Delta_2 = U_1 + U_3 - 2U_2$, introduced in Ref. \cite{koshino_gate-induced_2009}. These parameters fully characterize the distribution of the potential in the sample. We first note that the value of $\Delta_1$ is not zero but 0.25 eV in the absence of the gate voltage. Such a large potential asymmetry can be attributed to environmental impurities in contact with the sample, whose charging effect is not symmetrical with respect to the graphene plane. Since $\Delta_1$ depends on $V_g$, the gate also contributes to the total electric field across the sample. The physical meaning of the parameter $\Delta_2$, which is also finite at $V_g = 0$ and shows a noticeable gate voltage dependence, will become clear in the following.\\
Within a simple electrostatic model \cite{koshino_gate-induced_2009,avetisyan_electric_2009,mccann_electronic_2012} where each graphene layer is treated as an infinitely thin charged plane the electrostatic potentials can be straightforwardly related to the charge densities:

\beq
\Delta_1 = \frac{e^2 d}{\varepsilon_0 \varepsilon_r} (n_1 - n_3 ) + \Delta^{\rm ext}_1\, ,
\label{eq:delta1_expli}
\eeq

and

\beq
\Delta_2 = - \frac{e^2 d}{\varepsilon_0 \varepsilon_r} n_2\, .
\label{eq:delta2_expli}
\eeq

where $e$ is the elementary charge, $\varepsilon_0$ the vacuum permittivity, $d \approx 0.335 $ nm the interlayer distance and $\varepsilon_r$ the interplane permittivity between graphene layers due to the screening effect of all electrons except the $\pi$-band electrons. The quantity $\Delta^{\rm ext}_1$ takes into account the effect of the gate voltage and the environmental charges, which would produce a potential difference exactly equal to $\Delta^{\rm ext}_1$ over the distance of $2d$ in the absence of electrostatic screening by graphene.\\
Interestingly, all the quantities that enter Equation \ref{eq:delta2_expli}, except $\varepsilon_r$, are known from the infrared data, which allows us to derive the interplane permittivity. From Fig. \ref{fig:fitparam} we cannot identify any dependence of $\varepsilon_r$ on the gate voltage, and its value is $3.7 \pm 0.1$.\\
To our knowledge, there are no direct measurements of this quantity, which is however an important parameter needed to calculate the electronic screening in graphene. One should note that no relation analogous to Equation \ref{eq:delta2_expli} exists in the case of bilayer graphene allowing such a straightforward determination of $\varepsilon_r$. It turns out that the high doping level of our sample is beneficial for the accuracy of this method since $\varepsilon_r$ is proportional to the ratio of $n_2$ and $\Delta_2$, which would both be very small in an undoped sample.\\%
Once the value of $\varepsilon_r$ is determined, we can calculate $\Delta^{\rm ext}_1$ with Equation \ref{eq:delta1_expli} for each gate voltage (light gray squares in Fig. \ref{fig:fitparam}). $\Delta^{\rm ext}_1$ decreases linearly as a function of $V_g$ with a slope of about -0.0014 eV/V. Within the same electrostatic model, the slope is expected to be $\d \Delta^{\rm ext}_1 / \d V_g = -(d e \varepsilon_r)\cdot(\varepsilon_b/L_b) = -0.0012$ eV/V, where $\varepsilon_b$ = 3.9 and $L_b$ = 300 nm are the dielectric constant and the thickness of the oxide layer respectively. The experimental and the theoretical values are thus very close, supporting the model.

It is interesting to follow the modification of the band structure by the gate voltage. The bands corresponding to the best fits of the optical spectra are displayed for $V_g$ = -80 V and $V_g$ = +80 V in Fig. \ref{fig:bands}. Although the bands for the presented voltages are qualitatively similar, their positions and shapes, especially near the K point, are significantly influenced by the electric field. In both cases, there is a gap  of about 50 meV between the valence and the conduction bands. However, the chemical potential does not lie inside the gap, giving rise to a metallic, rather than semiconducting electronic properties. Notably in our highly doped sample the gap is much larger than the prediction for weakly doped ABA-stacked trilayers \cite{avetisyan_electric_2009,wu_field_2011,avetisyan_electric-field_2009}.\\
In conclusion, we investigated highly p-type doped ABA-stacked trilayer graphene with infrared spectroscopy. By fitting the experimental reflectivity spectra with a tight-binding model and the Kubo formula we extract the intrinsic parameters such as the carrier density and potentials of individual graphene layers. With a simple electrostatic model we can account for the effect of environmental charging and determine the interlayer permittivity between two graphene layers due to all the electronic shells except the $\pi$-bands. We show that, due to a large potential difference across the layers, a sizeable band gap in ABA-trilayer graphene is induced. Our work demonstrates new abilities of infrared spectroscopy in studying electronic screening in graphene multilayers.

\acknowledgments
We acknowledge I.Crassee and M.Fogler for useful discussions. We acknowledge the assistance of J. Teyssier for the Raman measurements. This research was supported by the Swiss National Science Foundation by the grants 200020-130093, 200020-140710 and IZ73Z0-128026 (SCOPES program), through the National Center of Competence in Research Materials with Novel Electronic Properties - MaNEP.%

\bibliographystyle{epj}
\bibliography{publi_zotero}

\end{document}